\documentstyle[12pt] {article}
\begin{document}
\thispagestyle{empty}

\title{ Neutrino Annihilation in Stellar Magnetic Fields}
\author{ C.J. Benesh and C.J. Horowitz,\\
Nuclear Theory Center, Indiana University,\\
Bloomington, IN 47405}

\maketitle
\setcounter{page}{0}

\begin{abstract}

	     The potential enhancement of the cross section 
$\nu\bar\nu\rightarrow e^+e^-$ in the presence of a magnetic field 
is of critical interest for the study of supernovae and as a possible 
mechanism for gamma ray bursts. While parity violation(PV) in this 
reaction in free space is forbidden by CP, the presence of a CP
non-invariant background gas of electrons creates an asymmetry in the cross
section which may contribute to the asymmetry of a supernova and the natal 
velocities of neutron stars.    
We calculate the cross section in the presence of fields as high as
$B = 10^{16}$G and find no significant enhancement of the cross section
with magnetic field strength.
By studying the systematics of our results as a function of environmental
variables(field strength, density,temperature), we extrapolate the
relative strength of the parity violating terms to the weak field limit,
and find the parity violation is insufficient to provide the ``kick''
required to explain the observed velocities of neutron stars.  

\end{abstract}

\pagebreak

\section{Introduction}

	Violent stellar events, such as supernovae and neutron 
star collisions, release tremendous amounts of energy, primarily 
in the form of neutrinos\cite{1}. In regions of low density, the 
annihilation process $\nu\bar\nu\rightarrow e^+e^-$
has been proposed as a source of energy to restart the shock wave in 
core collapse supernovae\cite{2}, and as a means of generating an energetic 
$e^+e^-$ plasma for gamma ray bursts\cite{3}. Since detailed 
calculations of supernova and gamma ray burst scenarios indicate that the
annihilation rate is too low\cite{4}, it has been suggested that the 
star's magnetic field will act as a catalyst for the reaction by eliminating 
some kinematical constraints on the $e^+e^-$ pair\cite{3}. The case for
such an enhancement is motivated by studies of pair creation
 ($\gamma\rightarrow e^+e^-$ and $\gamma\rightarrow \gamma\gamma$)\cite{5} 
 and the neutrino synchotron process 
($\nu\rightarrow\nu e^+e^-$\cite{6} \cite{7}), where processes 
that are kinematically disallowed in the vacuum are made possible by 
the exchange 
of momentum between the $e^+e^-$ pair and the magnetic field. The central role 
of the magnetic field in these processes is reflected in the sensitivity of the 
cross sections to the field strength. Significantly, this enhancement does not
seem to occur in processes that are kinematically allowed in the absence of the 
field\cite{8}.   
 
	As an added bonus, the magnetic field provides a preferred direction 
in space, opening the way for parity violating effects to produce an 
asymmetry in neutrino cross sections. The asymmetries, which have been calculated
for $\nu-e$ scattering\cite{7}\cite{9} and $\nu$-nucleus elastic 
scattering\cite{10}, provide a mechanism 
for producing asymmetric supernova explosions, either directly or by
asymmetrically heating the surrounding matter. Such asymmetric explosions have 
been proposed as a means of producing the observed large 
velocities of pulsars\cite{6}. 
	
	In this paper, we calculate the $\nu\bar\nu$ annihilation 
for a variety of magnetic field strengths, densities, and temperatures
similar to those found in supernovae. Our formalism\cite{8}\cite{9}\cite{10}, is 
detailed in the next section, while our results
and their implications for astrophysics are left to the concluding section.

\section{Formalism}  

	We consider the process for $\nu\bar\nu$ annihilation 
in the presence of a magnetic field of strength ${\bf B}$ in the z direction, 
using the four point interaction
\begin{equation}
L_{int} ={G_F\over\sqrt{2}}\, \int\, d^4x\, \bar\nu(x)\gamma_\alpha(1-\gamma_5)\nu(x)\,\bar e(x)\gamma^\alpha(C_V-C_A\gamma_5)e(x),
\end{equation}
where $G_F=1.13\times 10^{-11}$ MeV$^{-2}$ is the Fermi coupling strength,
$C_V=2\sin^2\theta_W\pm \frac{1}{2}$, $C_A=\pm\frac{1}{2}$ with the plus sign
for electrons and the minus for $\mu$ or $\tau$ neutrinos, and 
$\sin^2\theta_W\approx .223$.

	The wavefunctions of the electron-positron pair are obtained by 
solving the Dirac equation in Landau gauge ($A_x=-By$)\cite{12} 

\begin{eqnarray} 
\Psi^{e^-}_{(p_x,p_z,n,\sigma)}({\bf x},t)& =& {e^{ip_x x}e^{ip_z z}
e^{-i\epsilon t}\over \sqrt{L_xL_z}} \pmatrix{ \alpha CH_{n-1}(\xi)\cr
-\sigma \alpha DH_{n}(\xi)\cr \sigma\beta CH_{n-1}(\xi)\cr  -\beta DH_{n}(\xi)
\cr},\nonumber\\
\Psi^{e^+}_{(p^\prime_x,p^\prime_z,m,\sigma^\prime)}({\bf x},t)& =& 
{e^{-ip^\prime_x x}e^{-ip^\prime_z z}
e^{-i\epsilon^\prime t}\over \sqrt{L_xL_z}} \pmatrix{ \beta^\prime D^\prime 
H_{m-1}(\xi^\prime)\cr
-\sigma^\prime \beta^\prime C^\prime H_{m}(\xi^\prime)\cr -\sigma^\prime
\alpha^\prime 
D^\prime H_{m-1}(\xi^\prime)\cr  
\alpha^\prime C^\prime H_{m}(\xi^\prime)
\cr},
\end{eqnarray}

\noindent where $\pmatrix{\alpha\cr\beta}=\frac{1}{\sqrt{2}}
(1\pm\frac{m_e}{\epsilon})^{\frac{1}{2}}$, 
$\pmatrix{C\cr D}=\frac{1}{\sqrt{2}}(1
\pm{\sigma p_z\over (\epsilon^2-m^2)^{\frac{1}{2}}} )^{\frac{1}{2}}$,
$\,\sigma=\pm 1$ is the electron's helicity,
$H_n(\xi)$ is the normalized solution of a one dimensional harmonic 
oscillator with $\xi=\sqrt{eB}(y-{p_x\over eB})$, and the energies of the 
Landau levels are given by $\epsilon = \sqrt{p_z^2+2eBn+m_e^2}$.
For the positron spinor, the primed quantities are obtained from the
unprimed by replacing $\epsilon\rightarrow\epsilon^\prime$, $p_i\rightarrow
p^\prime_i$, and $\xi^\prime=\sqrt{eB}(y+{p^\prime_x\over eB})$.

	The differential cross section for the annihilation process is given 
by	
\begin{equation}
d\sigma = {G_F^2 L_xL_z\over 4\pi k_1\cdot k_2 V} C_{\mu\nu}A^{\mu\nu}
\delta(Q_0-\epsilon-\epsilon^\prime)\delta(Q_z-p_z-p^\prime_z)
\delta(Q_x-p_x-p^\prime_x) dp_x dp_z dp^\prime_x dp^\prime_z,
\end{equation}
where $Q^\mu=k_1^\mu+k_2^\mu$, with $k_1(k_2)$ the (anti-)neutrino momentum,
$V=L_xL_yL_z$ is a normalization volume,	
\begin{equation} 
C^{\mu\nu}= k_1^\mu k_2^\nu +k_1^\nu k_2^\mu - g^{\mu\nu}k_1\cdot k_2
+i\epsilon^{\mu\nu\alpha\beta}k_{1\alpha} k_{2\beta},
\end{equation}
and $A^{\mu\nu}=\sum_{\sigma\sigma^\prime} S^\mu S^{*\nu}$ is the 
spin-summed product of the current matrix elements of the electron.
Neglecting the electron mass, the matrix elements are given by
\begin{eqnarray}
S^0 &=& {(1-\sigma\sigma^\prime)\over 2}(C_V-C_A\sigma)
(-DC^\prime F_{nm}+CD^\prime F_{n-1,m-1})\nonumber\\
S^1 &=& {(1-\sigma\sigma^\prime)\over 2}(C_V-C_A\sigma)
(-DD^\prime e^{i\phi}F_{n,m-1}-CC^\prime e^{-i\phi}
F_{n-1,m})\nonumber\\
S^2 &=& -i{(1-\sigma\sigma^\prime)\over 2}(C_V-C_A\sigma)
(-DD^\prime e^{i\phi} F_{n,m-1}+CC^\prime e^{-i\phi}
F_{n-1,m})\nonumber\\
S^3 &=& {(1-\sigma\sigma^\prime)\over 2}(C_V-C_A\sigma)\sigma
(DC^\prime F_{nm}+CD^\prime F_{n-1,m-1}),
\end{eqnarray}
where
\begin{equation}
 F_{nm} = \left(\frac {n_<!}{n_>!}\right)^\frac{1}{2} \left(\frac{(n-m)}{\vert n-m\vert}\sqrt{\frac{
Q_\perp^2}{2eB}}\right)^{n_<-n_>} e^{-Q_\perp^2\over 4eB} e^{i(m-n)\phi} 
L_{n<}^{n_>-n_<}\left(\frac{Q_\perp^2}{2eB}\right),
\end{equation}
with $L_n^\alpha$ a generalized Laguerre polynomial\cite{A+S} and 
$\tan{\phi}=Q_y/Q_x$.  Here $n_>$ is the greater of $n$ and $m$.

	To simplify the remainder of the calculation, we choose coordinates 
such that $Q_y =0$. In this system, the non-zero components $A_{\mu\nu}$ are 
given by
\begin{eqnarray}
A_{00}&=&\frac{1}{2}\bigg[(C_V^2+C_A^2)\bigg(
(1+v_e v_p)(\vert F_{nm}\vert^2+
\vert F_{n-1,m-1}\vert^2) + 2 v_n v_m 
 F_{nm}F_{n-1,m-1}\vert_{\phi=0}\bigg) \nonumber\\
& & -2C_VC_A(v_e+v_p)\bigg(
\vert F_{n-1,m-1}\vert^2-\vert F_{nm}\vert^2
\bigg)\bigg]   \nonumber\\
& &\nonumber\\
A_{11}&=&\frac{1}{2}\bigg[(C_V^2+C_A^2)
\bigg((1-v_e v_p)(\vert F_{n,m-1}\vert^2+
\vert F_{n-1,m}\vert^2)-2 v_nv_m F_{n,m-1}
F_{n-1,m}\vert_{\phi=0}\bigg)\nonumber \\
& &  
+2C_VC_A(v_e-v_p)\bigg( \vert F_{n-1,m}
\vert^2-\vert F_{n,m-1}\vert^2 \bigg)\bigg]\nonumber\\
A_{22}&=&\frac{1}{2}\bigg[(C_V^2+C_A^2)\bigg((1-v_e v_p)
(\vert  F_{n,m-1}\vert^2+
\vert F_{n-1,m}\vert^2)+2 v_nv_m F_{n,m-1}F_{n-1,m}
\vert_{\phi=0}\bigg) \nonumber \\
& &+2C_VC_A(v_e-v_p)\bigg(
\vert F_{n-1,m}\vert^2-\vert F_{n,m-1}\vert^2
\bigg)\bigg]\nonumber\\
A_{33}& = & \frac{1}{2} \bigg[ 
(C_V^2+C_A^2) \bigg( (1+v_e v_p)(\vert F_{n,m}\vert^2+
\vert F_{n-1,m-1}\vert^2) - 2 v_n v_m  F_{n,m} 
F_{n-1,m-1} \vert_{\phi=0}\bigg) \nonumber\\
& & -2C_VC_A(v_e+v_p)( \vert F_{n-1,m-1}\vert^2-
\vert F_{n,m} \vert^2 
)\bigg]\nonumber\\
A_{01}&=& A_{10}=\nonumber\\
& &\,\frac{1}{2} \bigg[-(C_V^2+C_A^2)
\bigg( v_n (  F_{n,m-1}F_{n-1,m-1}+
 F_{n-1,m}F_{n,m}) +v_m ( F_{n,m-1}F_{n,m}
+ F_{n-1,m}F_{n-1,m-1})\bigg)
\nonumber\\
& &  +2C_VC_A \bigg( v_nv_p (F_{n,m-1}F_{n-1,m-1}
-  F_{n-1,m}F_{n,m})-v_mv_e (F_{n,m-1}F_{n,m}
- F_{n-1,m}F_{n-1,m-1})\bigg) \bigg]\vert_{\phi=0}
\nonumber\\
A_{03}&=&A_{30}=\nonumber\\
& &\,\frac{1}{2}\bigg[-(C_V^2+C_A^2)\bigg(
(v_e+v_p)(\vert F_{n,m}\vert^2+\vert F_{n-1,m-1}\vert^2)+ 2v_nv_m  F_{n,m}
F_{n-1,m-1}\vert_{\phi=0}\bigg)\nonumber\\
& &+2C_VC_A(1+v_ev_p)
\bigg(\vert F_{n-1,m-1}\vert^2-\vert F_{n,m}\vert^2\bigg)\bigg]
\nonumber\\
A_{13}&=&A_{31}=\nonumber\\
& &\,\frac{1}{2} \bigg[ ( C_V^2+C_A^2 )
\bigg(v_nv_p (F_{n,m-1} F_{n-1,m-1}
+ F_{n-1,m}F_{n,m}) +v_mv_e( F_{n,m-1}F_{n,m}
+ F_{n-1,m}F_{n-1,m-1})\bigg)\nonumber\\
& & -2C_VC_A \bigg( v_n (
 F_{n,m-1}F_{n-1,m-1}-
 F_{n-1,m}F_{n,m})+v_m(F_{n,m-1}F_{n,m}- F_{n-1,m}F_{n-1,m-1})\bigg)\bigg]
\vert_{\phi=0} 
\nonumber\\
A_{02}& = &-A_{20}=\nonumber\\
& &\,\frac{i}{2}
\bigg[ -(C_V^2+C_A^2)
\bigg( v_n(F_{n,m-1}F_{n-1,m-1}- F_{n-1,m}F_{n,m}) +v_m(F_{n,m-1}F_{n,m}
- F_{n-1,m}F_{n-1,m-1})\bigg) 
\nonumber\\
& & +2C_VC_A\bigg(v_nv_p( F_{n,m-1}
F_{n-1,m-1} + F_{n-1,m}F_{n,m})+v_mv_e( F_{n,m-1}F_{n,m}
+ F_{n-1,m}F_{n-1,m-1})\bigg)\bigg]\vert_{\phi=0}
\nonumber\\
A_{12}&=&-A_{21}=\nonumber\\
& &\, \frac{i}{2}\bigg[(C_V^2+C_A^2)(1-v_ev_p)
\bigg( \vert F_{n-1,m}\vert^2-\vert F_{n,m-1}\vert^2\bigg)\nonumber\\
& & +2C_VC_A\bigg((v_e-v_p)(\vert F_{n,m-1}
\vert^2+ \vert F_{n-1,m} \vert^2)-2 v_nv_m F_{n,m-1}
F_{n-1,m}\vert_{\phi=0}\bigg)\bigg] 
\nonumber \\
A_{23}&=&-A_{32}=\nonumber\\
& &\, \frac{1}{2} \bigg[( C_V^2+C_A^2)
\bigg(v_nv_p(F_{n,m-1}F_{n-1,m-1}- F_{n-1,m}F_{n,m})
 -v_mv_e (F_{n,m-1}F_{n,m} -F_{n-1,m}F_{n-1,m-1})\bigg) \nonumber\\
& &
 +2C_VC_A \bigg( v_n ( F_{n,m-1}
F_{n-1,m-1}- F_{n-1,m}F_{n,m})-v_m (  F_{n,m-1}F_{n,m}
- F_{n-1,m}F_{n-1,m-1})\bigg)\bigg]\vert_{\phi=0},
\end{eqnarray}
\noindent where
$v_{n(m)}=\sqrt{2eBn(m)}$,$v_{e}=\frac{p_z}{\epsilon}$,and $v_p=
\frac{p^\prime_z}{\epsilon^\prime}$. The contributions may be classified as
either parity conserving($\propto (C_V^2+C_A^2)$, symmetric) or 
parity violating, and by whether the parity violation occurs 
in the neutrino($\propto (C_V^2+C_A^2)$, anti-symmetric) or electron currents
($\propto C_VC_A$, symmetric). This last distinction is important 
both because the total effect of parity violating neutrino currents will
tend to vanish when integrated against the initial spectrum of neutrino and 
anti-neutrinos, and because the asymmetric propagation of neutrinos vs 
anti-neutrinos will lead to asymmetries in the the star's lepton number,
$Y_e$.    

	All that remains is to solve the kinematic constraints on the
electron momenta and to include medium effects. The former is
a straight forward algebraic exercise, yielding
\begin{equation}
p_z={(Q_0^2-Q_z^2+2eB(n-m))Q_z+\lambda Q_0\sqrt{(Q_0^2-Q_z^2)^2-
4eB(Q_0^2-Q_z^2)(n+m)+4e^2B^2(n-m)^2}\over 2(Q_0^2-Q_z^2)},
\end{equation}
 with $p^\prime_z=Q_z-p_z$ and $\lambda=\pm1$. Implicit in this 
prescription are
 maximum values for $m$ and $n$, $N_{max}=int((Q_0^2-Q_z^2)/2eB)$,
and $M_{max}= int{(\sqrt{N_{max}}-\sqrt{n})^2)}$. Additionally, the
integration over the energy conserving delta function will generate 
a Jacobian factor $\vert v_e-v_p\vert^{-1}$. After inserting Pauli 
blocking factors for the electron and positron, the cross section becomes
\begin{equation}
\label{finito}
\sigma= \sum_{\lambda=\pm 1}\sum_{n=0}^{N_{max}}\sum_{m=0}^{M_{max}}
{G_F^2 eB\over 2\pi s} C_{\mu\nu}A^{\mu\nu} 
{(1-f_e)(1-f_p)\over\vert v_e-v_p\vert}
\end{equation}
where $s=(k_1+k_2)^2$ and 
\begin{eqnarray}
f_{e}&=&{1\over 1+e^{{\epsilon-\mu_0}\over T}},\nonumber\\
f_{p}&=&{1\over 1+e^{{\epsilon^\prime+\mu_0}\over T}}
\end{eqnarray}
with $\mu_0$ the electron chemical potential and $T$ the temperature
of the medium.

\section{Results}

	In order to understand the behaviour of the cross section
derived in the last section, we have performed calculations for two 
different kinematical situations using a variety of field strengths,
temperatures and densities typical of those found in astrophysical 
events. In the first of these, which we shall refer to as the collinear
case, the neutrino and anti-neutrino are traveling nearly parallel to
one another, as is appropriate for neutrinos escaping a supernova. We 
assume their four momenta are given by
\begin{eqnarray}
k_1 &=& (E,P\sin\theta,\sqrt{s}/2,P\cos\theta) \nonumber\\
k_2 &=& (E,P\sin\theta,-\sqrt{s}/2,P\cos\theta) 
\end{eqnarray} 
with $E=10 MeV$, and $P=9 MeV$. The second kinematical situation 
we shall consider is the case where the neutrino and anti-neutrino
collide head-on with zero center of mass momentum. In this instance,
the neutrino and anti-neutrino momenta are given by
\begin{eqnarray}
k_1 &=& (P,P\sin\theta,0,P\cos\theta)\nonumber\\
k_2 &=& (P,-P\sin\theta,0,-P\cos\theta).
\end{eqnarray}      
This situation, which desribes the annihilation of backscattered
(anti-)neutrinos by outgoing (anti-)neutrinos, represents the largest 
cross section for a given incident energy. For ease of comparison
with the symmetric case, we  choose $P=8.72$ MeV so that the center
of mass energy, and consequently the zero field cross section,
remains the same as in the case of parallel kinematics. 

	In both kinematical regimes, the number of possible final
states grows inversely with $B^2$, so that it is necessary to calculate 
the electron-positron matrix elements for large values of $n$ and $m$ in
order to obtain the cross section. In the case of head-on collisions,
this difficulty is mitigated by the fact that only matrix elements 
with $n=m$ are non-zero. For the symmetric collisions, however,
it is necessary to calculate Laguerre polynomials of large order.
This is accomplished by upward recursion using quadruple precision
arithmetic to avoid numerical instabilities. As an added precaution,
the resulting polynomials have been compared to results obtained from a 
downward recursive Clenshaw scheme, and found to agree to six significant
figures. Thus, our calculations are not limited by the accuracy to which
the Laguerre functions can be realized.  

	In Fig. 1, the cross section for $\nu\bar\nu\rightarrow e^+e^-$
in free space is shown for a variety of magnetic field strengths as a 
function of the angle between the total momentum and the magnetic field 
direction, assuming symmetric kinematics for the 
annihilating neutrinos. The jagged structure of the cross section
as a function of angle is a result of the near-vanishing of the
Jacobian factor at threshold values of $\cos\theta$ where new final
states for the electron-positron pair become available. For strong
fields($B=100 m_e^2(\approx 4\times 10^{16}$G)), this cross section 
varies strongly with angle, but this effect essentially vanishes
for fields comparable to those expected in the core of superova
($B/m_e^2 < 1)$. Qualitatively, the angle averaged cross section tends 
to decrease with increasing $B$, which reflects the fact that the creation 
of both the electron and positron in the lowest Landau level is forbidden 
by helicity conservation. Additionally, there is no asymmetry with respect to 
the magnetic field direction, as such a preference would violate CP. 
In Fig. 2, the cross section is shown for the head-on collisions for the same
center of mass energy. Because the electron and positron are only sensitive
to the total momentum of the annihilating pair, and to the direction of the 
B-field, the cross section varies quadratically with the angle between the 
field and ${\bf k}_1$, and can be shown analytically to have the form
\begin{equation}
\sigma\propto  (A_{11}+A_{33}) + \cos^2\theta (A_{11}-A_{33})
\end{equation}
For large magnetic fields, the $A_{11}$ term dominates, and the 
annihilation cross section for neutrinos traveling parallel to
the magnetic field is twice that for those traveling perpendicular to the 
field. As the field decreases, the electron-positron current matrix elements
are more isotropic, and the angular dependence of the cross section is
correspondingly smaller. Interestingly, there is a region at small fields
where $A_{33}>A_{11}$, so that neutrinos traveling perpendicular to the 
field are more likely to annihilate.  
 
 When the neutrinos annihilate in the presence of a nearly degenerate gas 
of electron-positron pairs typical of the astrophysical environments, the 
cross section is reduced significantly by the requirement that the produced
electron's energy be above the Fermi energy of the gas. Moreover, since 
the electron gas is not CP symmetric, the argument forbidding a parity 
violating asymmetry in the annihilation cross section is invalid. In Figs. 
3 and 4, the cross sections for collinear and head on collisions are 
shown as a function of magnetic field strength, assuming an electron
gas with $T=1$ MeV and $\mu=15$ MeV. Comparison of Figs. 1 and 3 
indicates that, at the energies shown, the cross section is suppressed by a 
factor of 5-10, and that the cross section for annihilating pairs with 
momentum parallel to the magnetic field direction is larger. The 
suppression of the cross section for head on collisions is much more 
dramatic, on the order of $10^3$ for the center of mass energies shown. 
The cross section is also observed to be more sensitive to magnetic field 
strength. This 
reflects the difficulty of producing a high energy electron in the
head on collision relative to the collinear case, where the electron can 
be chosen to carry the bulk of the neutrinos' initial momentum. In Figs. 5 
and 6, the neutrino annihilation asymmetry, defined as the ratio of the 
parity violating to parity conserving contributions to the cross section,
as a function of magnetic field strength and direction. In the collinear case, 
the asymmetry is, for angles not too near $0$ or $\pi$, well fit by a 
straight line in $\cos\theta$ with slope proportional to the strength of 
the magnetic field. For the head on case, the linear dependence on 
$\cos\theta$ can be demonstrated analytically since the electon-positron
currents depend only on the total momentum perpendicular to the magnetic field.
The resulting asymmetry is given by
\begin{equation}
R=\frac{\sigma_{PV}}{\sigma_{PC}}\propto\frac{A_{12}\cos\theta}{(A_{11}+A_{33}) + \cos^2\theta (A_{11}-A_{33})},
\end{equation}
with the same linear dependence on the field strength.

	As we have already mentioned, the asymmetry is allowed because the 
background gas of electrons and positrons is not CP symmetric. Since the
energies of the electrons in the magnetic field are spin dependent, there is
a corresponding spin dependence in the occupation probabilities for states
in a given Landau level. This variation in the occupation probabilities 
of states near the Fermi surface results in a preference for creating
electrons with spin aligned along the field direction. In Figs. 7 and
8, the ratio of the parity violating to parity conserving contributions
to the annihilation cross section is shown as a function of angle and
electron chemical potential $\mu$ for $B=10m_e^2$. For the collinear case
(Fig. 7), the asymmetry increases with $\mu$ approximately quadratically,
reflecting the increase in the density of states near the Fermi surface.
For the head-on case(Fig. 8), the asymmetry shows no significant dependence 
on $\mu$ at all. This remarkable result comes about as a result 
of the parity violation residing in the neutrino matrix elements, which are
oblivious to the presence of the Fermi sea and to the direction of the
magnetic field. Memory of the field direction is recovered when the 
neutrino matrix elements are multiplied by the electron currents, producing
the observed asymmetry. The effect of increasing the temperature from
1-4 MeV is shown in in Figs. 9 and 10 for the same magnetic field strength.
As the Fermi surface becomes more diffuse, the electron occupation
numbers start to vary more slowly, with the result that the asymmetry
decreases for large $T$. In the collinear case(Fig. 9), the asymmetry falls
approximately like $\frac{1}{\sqrt{T}}$ while in the head-on kinematics the
fall-off is faster, varying approximately as $\frac{1}{T}$.  

	Combining these results, we are able to characterize the asymmetry
for the two cases in a manner that allows us to extrapolate our results
to other situations. For the collinear case, we find
\begin{equation} 
{\sigma_{PV}\over\sigma_{PC}}\propto {\mu^2 B\over\sqrt{T}}. 
\end{equation}
Since in this case there are two independent parameters with dimensions
of energy, $E$ and $s$, we have been unable to discover a simple
energy dependence for the asymmetry. For the head-on case, there is
only one dimensionally consistent possibility, given by
\begin{equation}
{\sigma_{PV}\over\sigma_{PC}}\propto {B\over T\sqrt{s}}.
\end{equation}
The dependence of the head-on cross section as a function of energy and angle
is shown in Fig. 11.

As we have noted previously, these relations are {\it approximate},
good to 10-15 per cent over a wide range of angles. Moreover, it should
be apparent that these relations are only valid in instances were the
asymmetry is fairly small, less than or of the order of 20 per cent. 
If we assume the validity of these relations, we can extrapolate from
the relatively high fields($\approx 10^{14}$ G) where calculations are 
feasible,
to the field strengths observed in pulsars($\approx 10^{12}$G)\cite{13}.  
The result of this scaling is an asymmetry of order $10^{-4}$, which is 
significantly smaller than that required to produce the observed recoil 
velocities of neutron stars. The smallness of this result does not, however, 
rule out neutrino asymmetries as a source of the recoil velocities since 
the relatively small asymmetries produced by neutrino annihilations may be 
amplified by hydrodynamic instabilities in the collapsing core\cite{14}. 
Additionally, there have been speculations that much larger magnetic fields 
may exist\cite{15}\cite{16}.  Finally, it should be noted that there are
other neutrino processes for which the cross section and asymmetries may be
larger, and multiple neutrino interactions may significantly increase the net 
asymmetry\cite{17}.   

	In summary, we have calculated the cross section for $\nu\bar\nu
\rightarrow e^+e^-$ in the presence of a magnetic field for an extensive
set of conditions typical of those found in astrophysical environments. 
We do not find that the presence of the field provides any significant
enhancement of the cross section at fixed chemical potential. Although 
the presence of a degenerate gas of electrons provides the possibility
of parity violation in this reaction, they are insufficient to produce 
the observed recoil velocities of neutron stars at the field strengths 
known to be associated with pulsars.  

\centerline{\bf Acknowledgments}

	This authors wish to acknowledge support from  NSF Research 
Contract No. PHY-9408843 and DOE grant DE-FG02-87ER40365.

\pagebreak
\begin{centering}
\centerline{\underline{Figure Captions}}
\end{centering}
\begin{itemize}
\item{} Fig. 1 - Neutrino annihilation cross section in free space for 
collinear kinematics as a function of magnetic field strength(in units of 
$m_e^2$) and direction of the neutrino pair's total momentum. 
\item{} Fig. 2 - Neutrino annihilation cross section in free space for
head-on collisions as a function of magnetic field strength(in units of
$m_e^2$) and direction of the neutrino's momentum. 
\item{} Fig. 3 - Neutrino annihilation cross section(collinear kinematics) at 
astrophysically interesting temperature($T=1$ MeV) and density($\mu=15$ MeV) 
as a function of magnetic field strength(in units of $m_e^2$)
and direction of the neutrino pair's total momentum.
\item{} Fig. 4 - Neutrino annihilation cross section(head-on kinematics) at 
astrophysically interesting temperature($T=1$ MeV) and density($\mu=15$ MeV) 
as a function of magnetic field strength(in units of $m_e^2$)
and direction of the neutrino's momentum.
\item{} Fig. 5 - Neutrino annihilation asymmetry for the collinear case
as a function of magnetic field strength and the direction of the neutrino 
pair's total momentum. The electron temperature and density are the same as
in Fig. 4.
\item{} Fig. 6 - Neutrino annihilation asymmetry for head on collisions
as a function of magnetic field strength and the neutrino's momentum. The 
electron temperature and density are the same as in Fig. 4.
\item{} Fig. 7 -  Neutrino annihilation asymmetry for the collinear case
as a function of chemical potential and the direction of the neutrino 
pair's total momentum. The electron temperature is the same as in Fig. 4,
and the magnetic field strength is $B=10m_e^2$.
\item{} Fig. 8 -  Neutrino annihilation asymmetry for head on collisions
as a function of chemical potential and the direction of the neutrino's 
momentum. The electron temperature is the same as in Fig. 4,
and the magnetic field strength is $B=10m_e^2$.
\item{} Fig. 9 -  Neutrino annihilation asymmetry for the collinear case
as a function of temperature and the direction of the neutrino 
pair's total momentum. The electron chemical potential the same as in Fig. 4,
and the magnetic field strength is $B=10m_e^2$.
\item{} Fig. 10 -  Neutrino annihilation asymmetry for head on collisions
as a function of temperature and the direction of the neutrino's 
momentum. The electron chemical potential is the same as in 
Fig. 4, and the magnetic field strength is $B=10m_e^2$.
\item{} Fig. 11 -  Neutrino annihilation asymmetry for head on collisions
as a function of neutrino center of mass energy and the direction of the 
neutrino's momentum. The electron chemical potential and temperature are 
the same as in Fig. 4, and the magnetic field strength is $B=10m_e^2$.
\end{itemize}
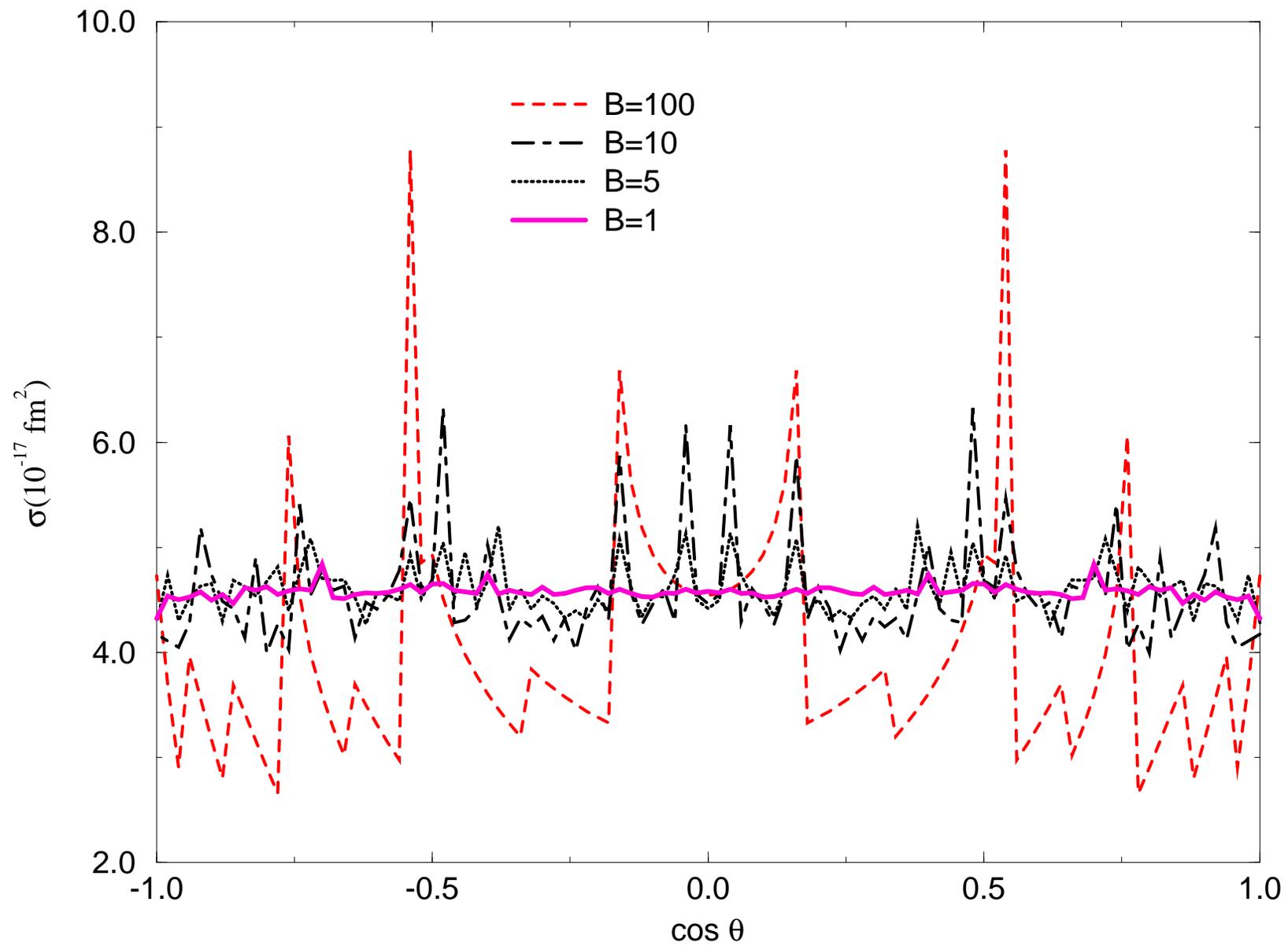
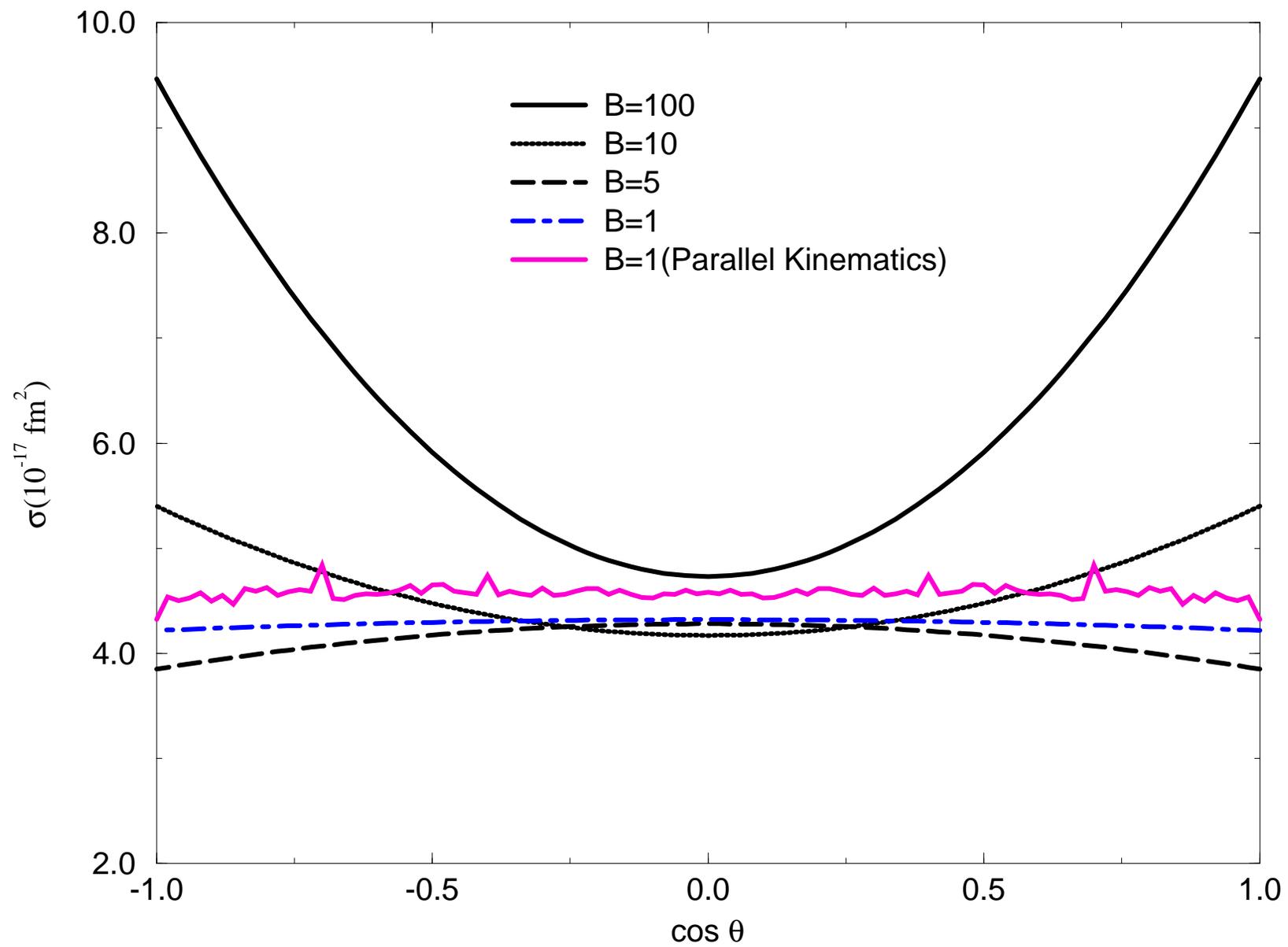
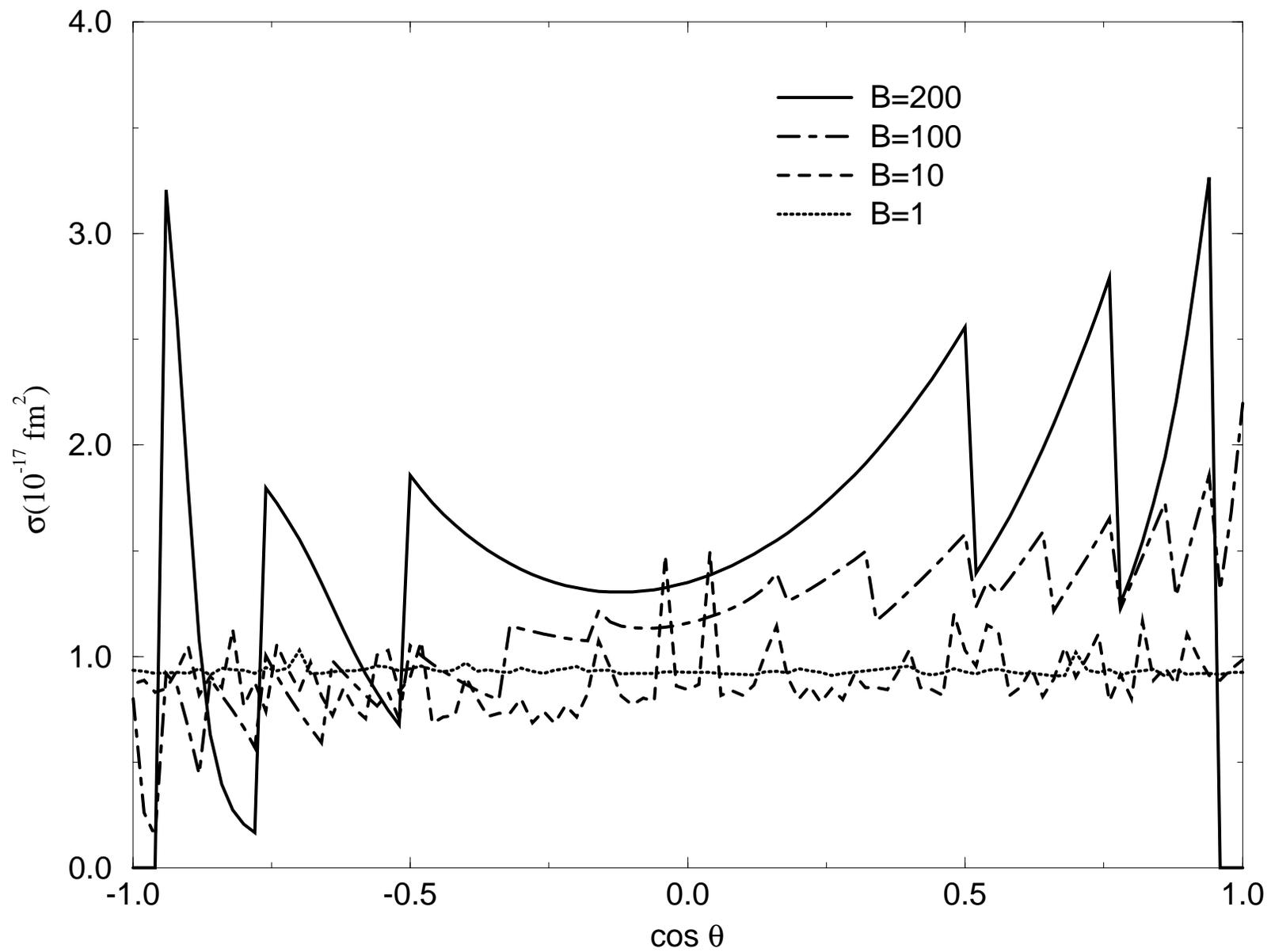
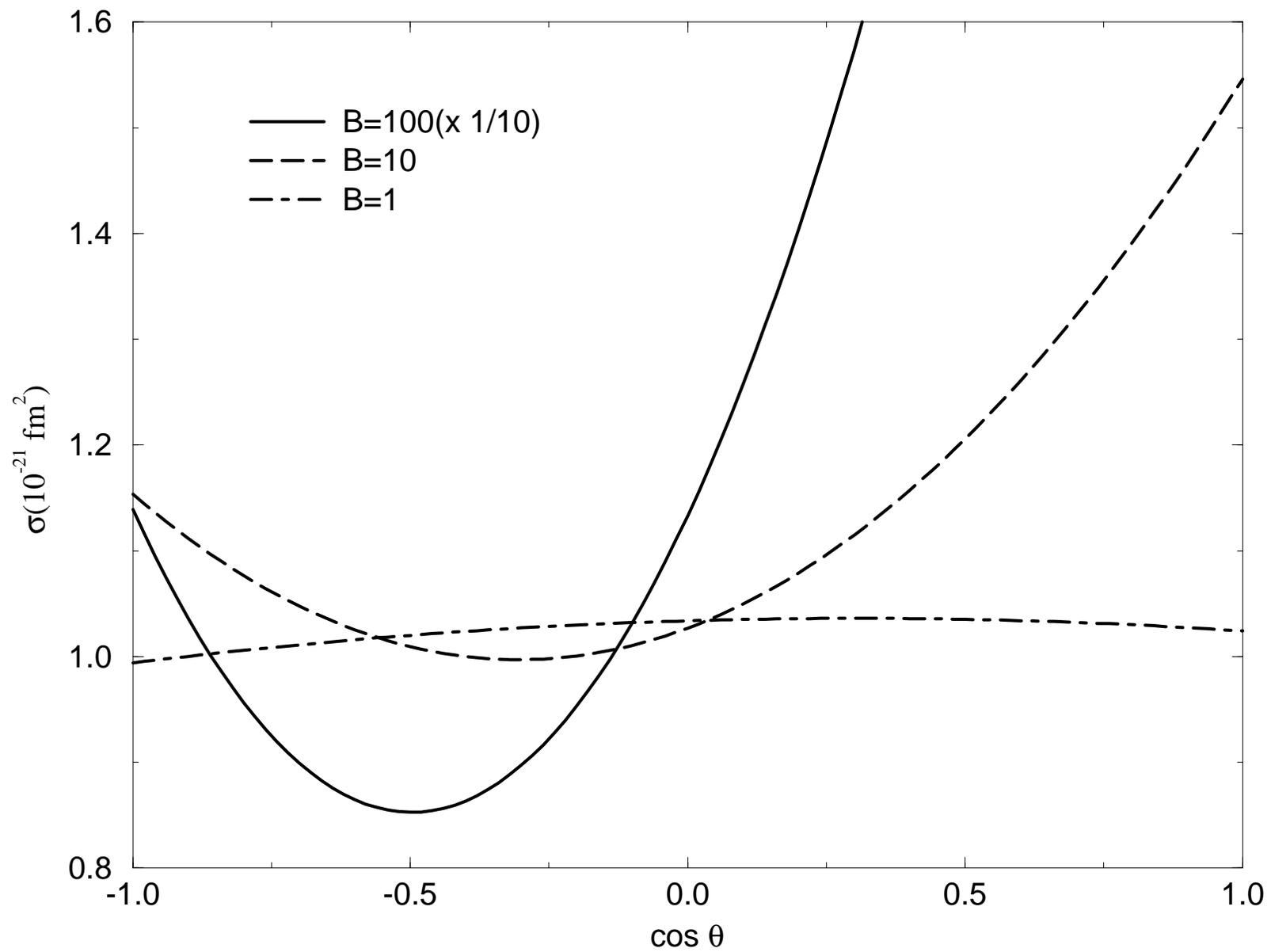
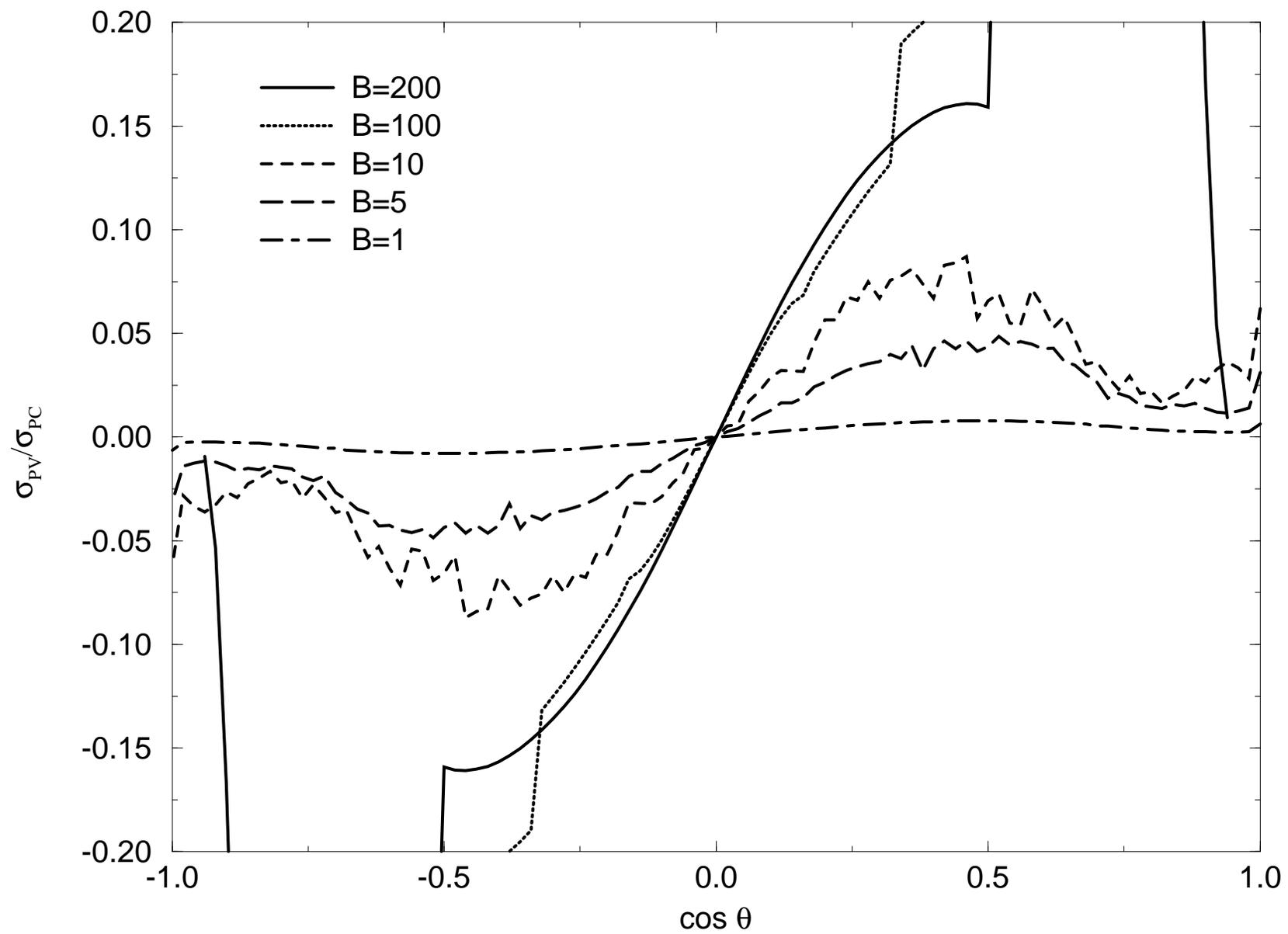
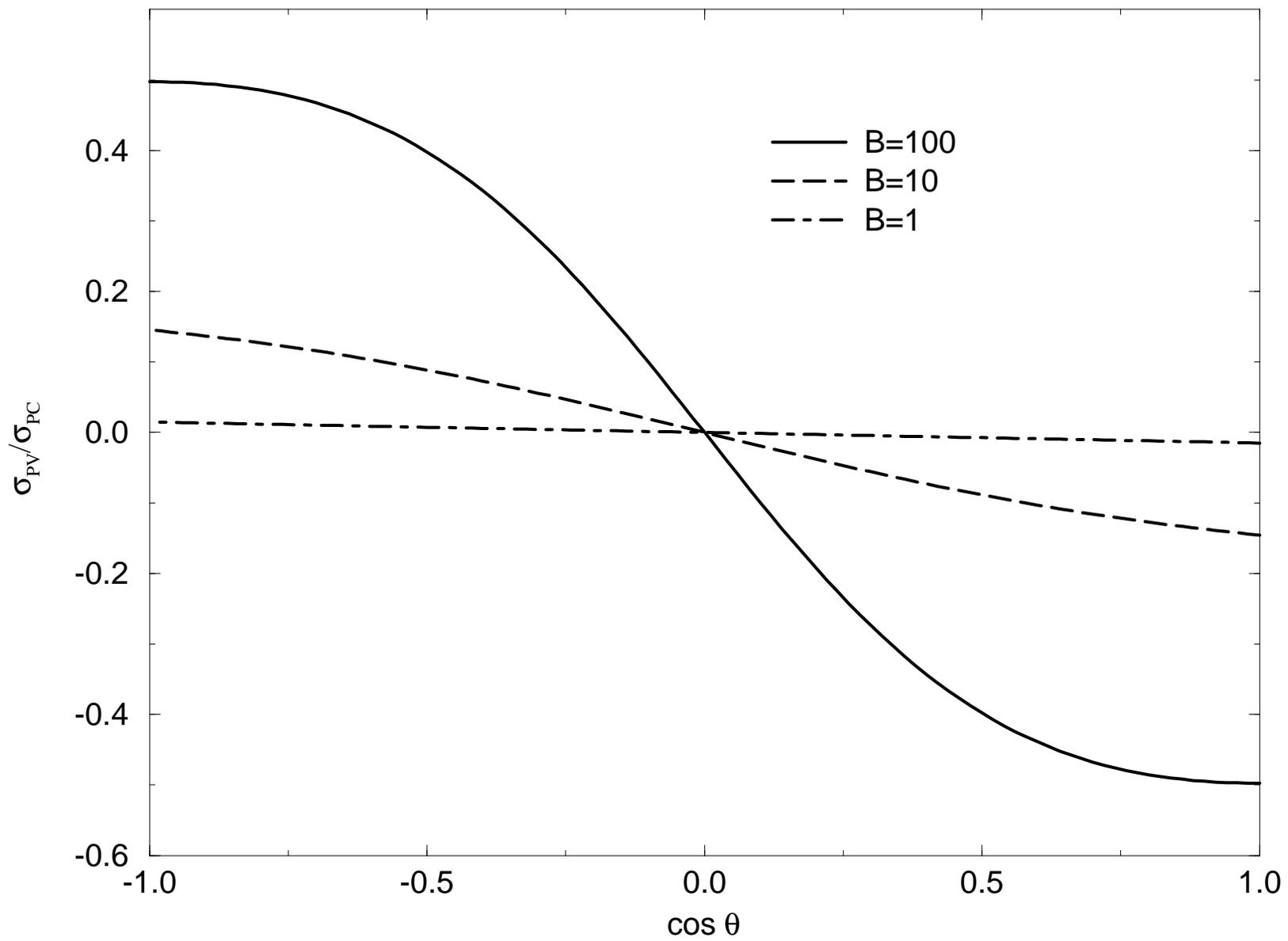
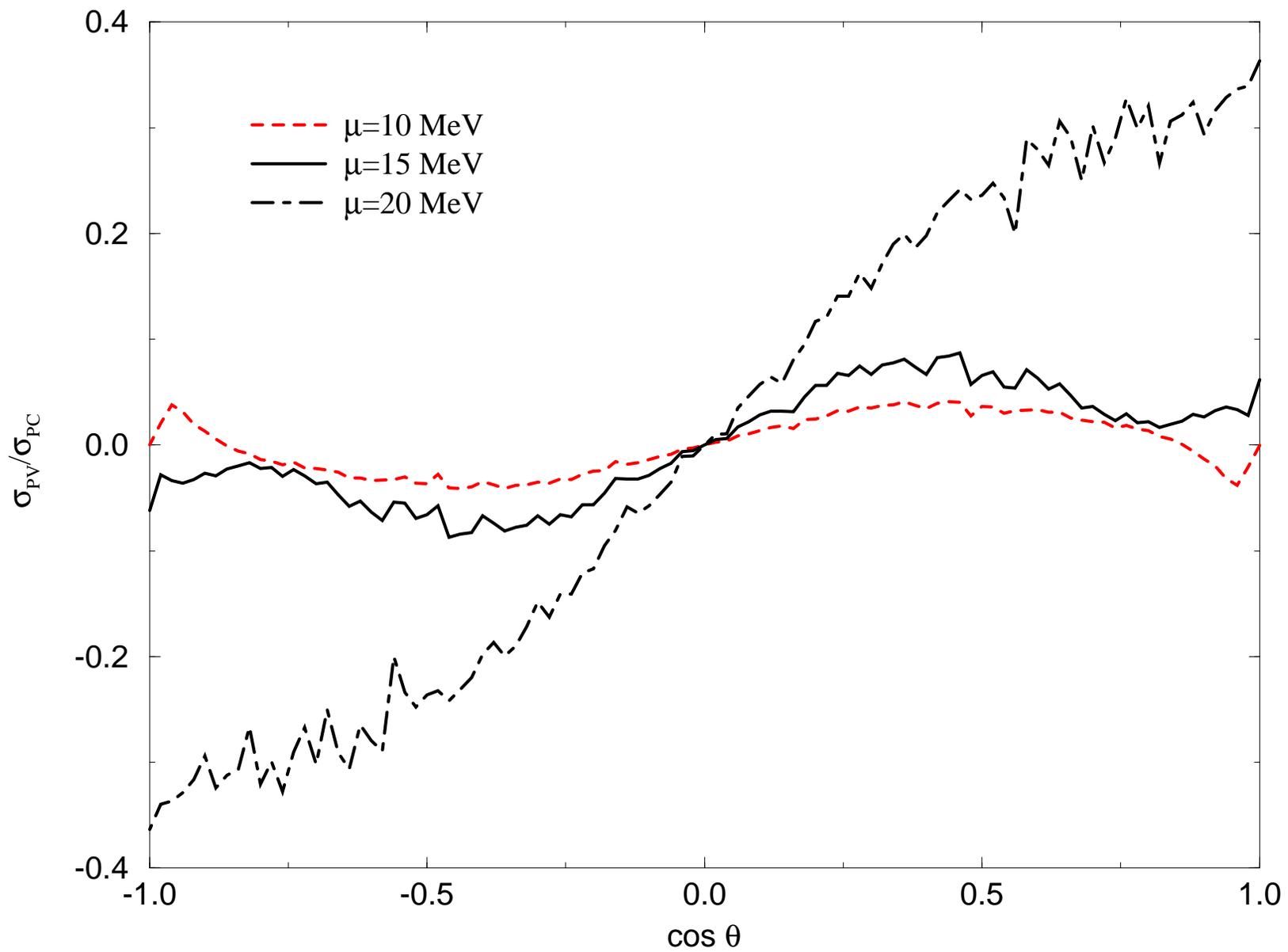
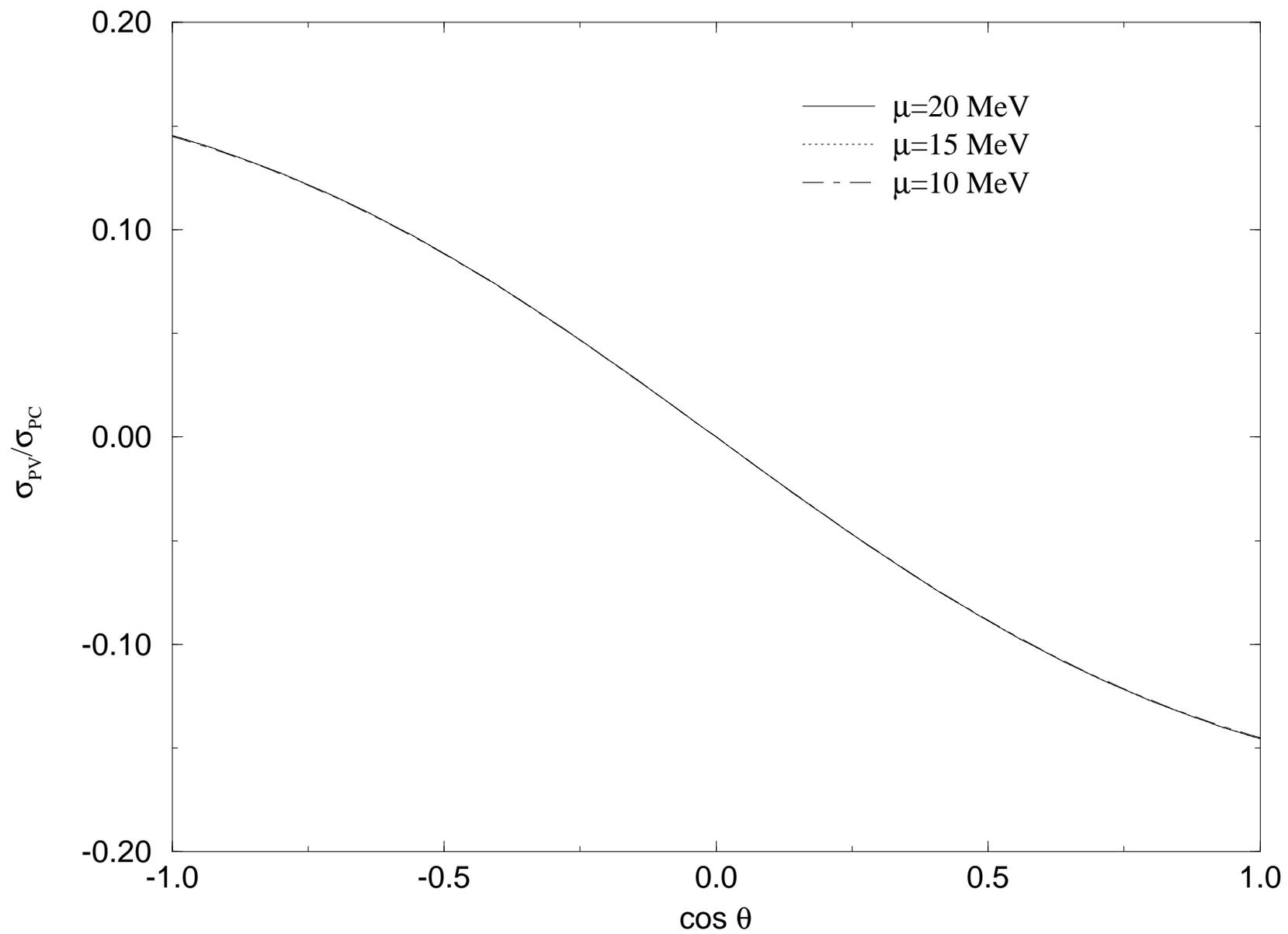
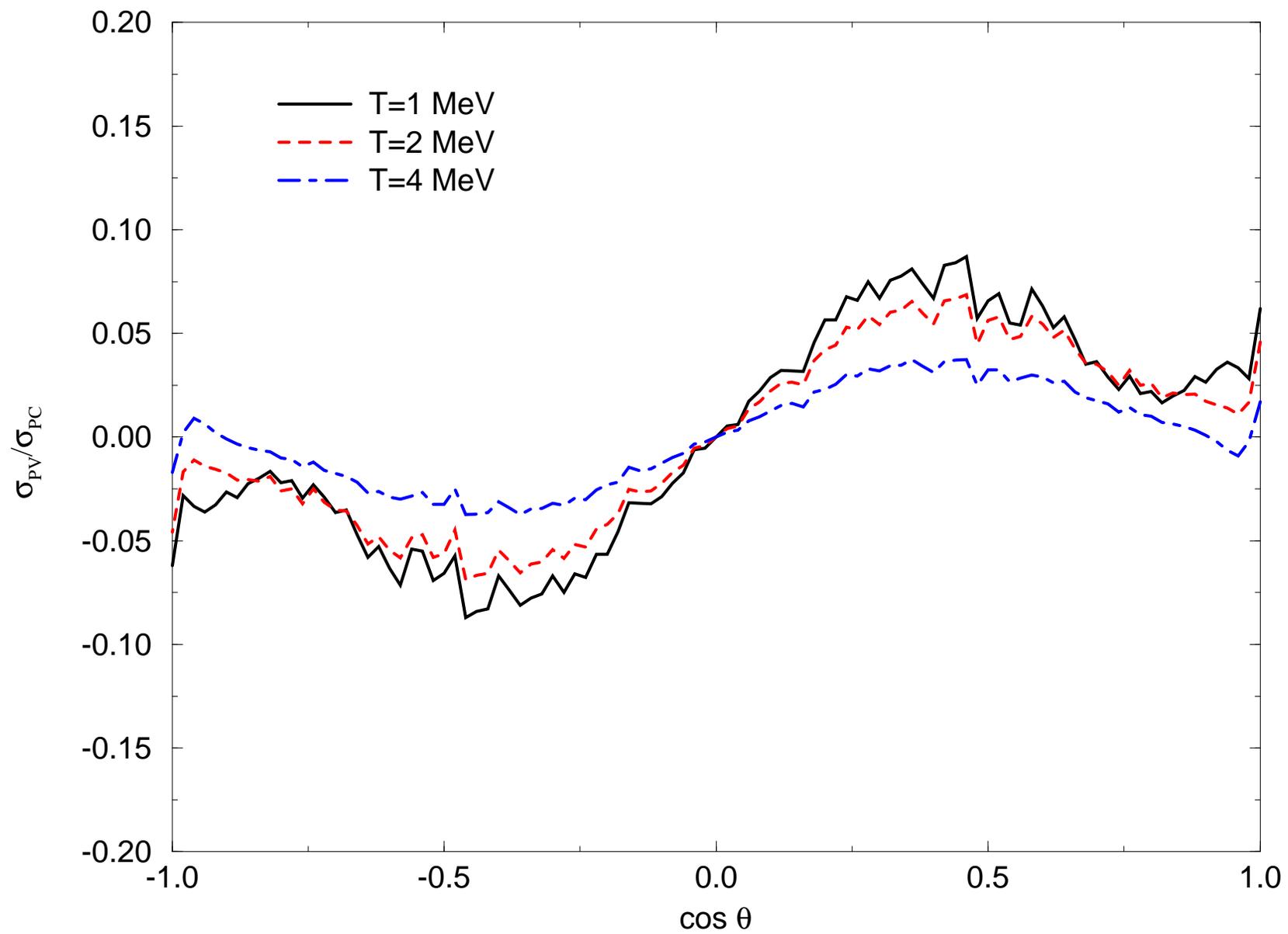
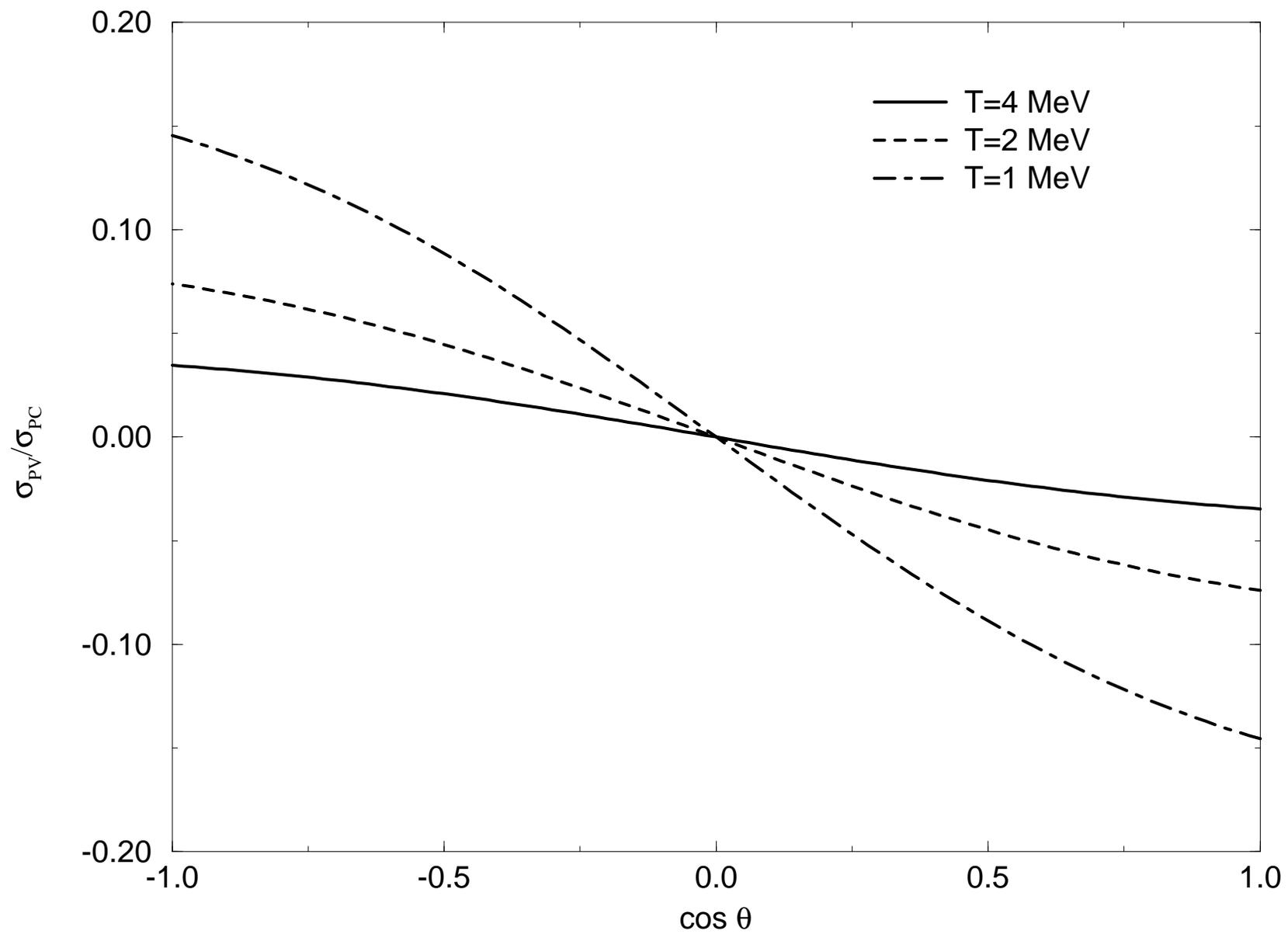
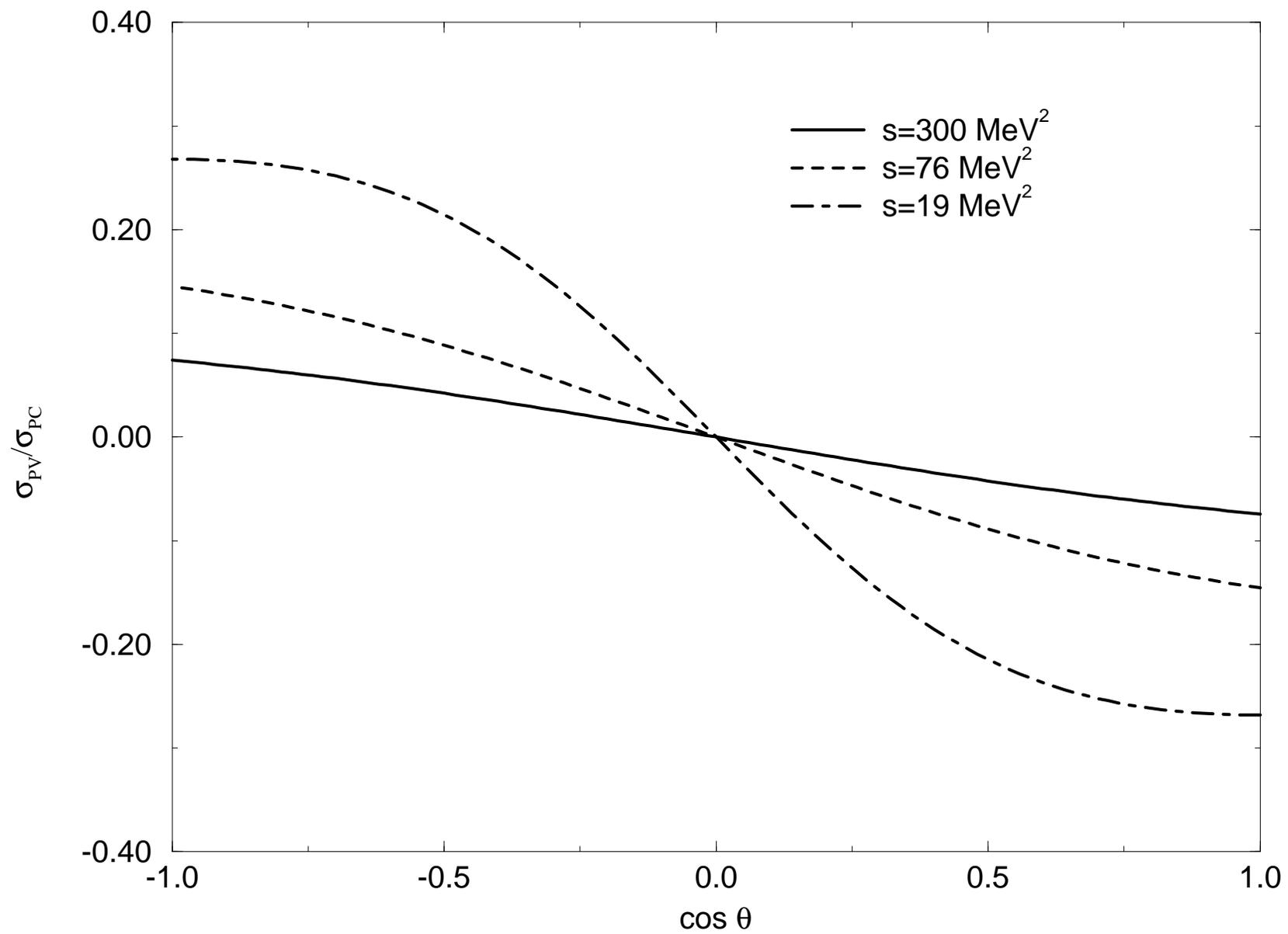
\end{document}